# The Z CamPaign: Year 1

**Mike Simonsen**
*AAVSO, 49 Bay State Road, Cambridge, MA, 02138; mikesimonsen@aavso.org*

**Abstract** The Cataclysmic Variable Section of the American Association of Variable Star Observers (AAVSO) has initiated an observing campaign to study a subset of dwarf novae (DNe), known as Z Cam type (UGZ). We call this program the Z CamPaign. Since there is no strong agreement between the various published catalogues as to which few dozen DNe are actually Z Cam type systems, our primary goal is to accumulate enough data to construct detailed light curves, covering the entire range of variability, to determine unequivocally the 30 Z CamPaign subjects' membership in the UGZ class of DNe. We discuss the organization, science goals, and some early results of the Z CamPaign in detail.

**1. Defining Z Cams**

U Geminorum-type (UG) variables, also called dwarf novae, are close binary systems in which a dwarf or subgiant K–M star fills the volume of its inner Roche lobe and loses mass to a white dwarf surrounded by an accretion disk. From time to time the system goes into outburst, brightening rapidly by several magnitudes. After several days to a month, or more, it returns to its original state. These dwarf nova outbursts are believed to be caused by thermal instabilities in the disk. Gas accumulates in the disk until it heats up and becomes viscous. This increased viscosity causes it to migrate in toward the white dwarf, heating up even more, eventually causing an outburst (Warner 1995).

Intervals between two consecutive outbursts for a given star may vary, but every star is characterized by a characteristic mean value of these intervals. This mean cycle corresponds to the mean amplitude of the outbursts. Generally speaking, the longer the cycle, the greater the amplitude of the outbursts. According to the characteristics of their light curves, UGs are further subdivided into three types: SS Cyg, SU UMa, and Z Cam. SU UMa type UGs are not relevant to this discussion.

SS Cygni-type variables (UGSS) increase in brightness by 2 to 6 magnitudes in *V* in 1 to 2 days and after several subsequent days return to their original brightness. The cycle times vary considerably, from 10 to several hundreds of days.

Z Camelopardalis-type stars (UGZ) also show cyclic outbursts, but differ from UGSS variables by the fact that sometimes after an outburst they do not return to their quiescent magnitude. Instead they appear to get stuck, for months or even years, at a brightness of about one magnitude fainter than outburst maximum.

These episodes are known as standstills. Z Cam cycle times characteristically range from 10 to 40 days, and their outburst amplitudes are from 2 to 5 magnitudes in *V*, but *standstills are the defining characteristic of the Z Cam stars.*

## 2. Standstills

If a dwarf nova has a high enough mass-transfer rate, it can resemble a dwarf nova continuously stuck in outburst. This is what nova-like variables are thought to be. One theory explaining Z Cam standstills is that the rate of mass transfer is approximately equal to the critical rate that separates dwarf novae from the nova-like variables (Meyer and Meyer-Hofmeister 1983).

Models can now explain why standstills are about a magnitude fainter than outburst maximum. The gas stream from the mass-losing star heats the disk, and because of this extra source of heat, the critical mass transfer rate at which a standstill occurs is about 40% less than the mass transfer rate during outburst (Stehle *et al.* 2001).

Even with the observations and analyses of recent years, we still know relatively little about standstills. Even the fundamental observational properties of how often they occur or how long they last are not well known. Standstills of Z Cam, the prototypical star of this class, can last between 9 and 1,020 days (Oppenheimer *et al.* 1998). Z Cam was in standstill almost continuously between 1977 and 1981 (AAVSO data). AH Her has been in standstill since June 2009, nine months and counting as of this writing. In contrast, HX Peg has much shorter standstills, from 30 to 90 days long, which can recur yearly (AAVSO data).

Standstills are not static affairs. Szkody and Mattei (1984) showed erratic flare-ups with amplitudes of several tenths of a magnitude in their compilation of the statistics of dwarf nova outbursts. Another well-quoted characteristic of Z Cams is that "standstills are always initiated by an outburst," and "standstills always end with a decline to quiescence" (Hellier 2001). However, there are at least five Z Cam stars that appear to go into outburst from standstill: V513 Cas, IW And, HX Peg, AH Her, and AT Cnc. If this turns out to be an inconvenient truth it provides another challenge in explaining the mechanisms that initiate and end standstills.

## 3. Science Goals

The science goals of the Z CamPaign are:

1. To determine convincingly which CVs are indeed UGZ and which are imposters. The plan is to analyze the light curves of all the candidates looking for standstill episodes in their light curves. If there are standstills, we will accept them as UGZ. If there are no standstills we will remove them from the list of known Z Cams, and assign another type to them, if possible.

If the data are inconclusive, we will concentrate on obtaining adequate long-term data throughout the range of the variable to make a determination.

2. To improve the overall data available on each of these stars and fill the gaps in the light curves. Since so little is known, even about the well observed Z Cam candidates, we will try to obtain as complete coverage as possible, concentrating on V magnitude observations first, then extending to other bandpasses.

3. To determine if some UGZ actually do go into outburst from standstill, or if perhaps we have just missed the sudden drop to quiescence before the next outburst, leading to the appearance of outburst from standstill behavior.

4. To study and report any other serendipitous discoveries about "UGZ-ness" that come to light as a result of improved coverage.

5. To publish the results in a peer-reviewed journal such as *The Journal of the AAVSO*.

**4. Coordinating the campaign**

The campaign is coordinated through the Cataclysmic Variable Section of the AAVSO. There is a special campaign page online explaining the details to those interested in observing these stars: https://sites.google.com/site/aavsocvsection/z-campaign. The list of campaign stars can be downloaded in several formats from this web page. The star list is divided into four sub-categories, based on the type and magnitude range.

The first group of stars are confirmed UGZ suitable for continued observation by visual observers throughout their cycles. These are generally the brightest dwarf novae in the campaign and have well sampled light curves, some going back as far as the 1940s. We strongly urge visual observers to continue monitoring these stars for their expected outbursts and standstills as well as unexpected behavior.

The second group of stars are unconfirmed UGZ stars that visual observers should continue to monitor for outbursts and standstills if or when they may occur.

The third group of stars are unconfirmed UGZs that both visual and CCD observers are encouraged to monitor for outbursts, but the standstills are likely to only be visible to CCD observers due to their relative faintness (15th or 16th magnitude). We encourage CCD observers to concentrate on these stars when they are known to be in outburst in particular, so they can monitor the fade from maximum looking for a standstill.

The last group are stars best suited to CCD observers for monitoring for outbursts and standstill behavior. These stars are too faint, even at maximum, for most visual observers to waste valuable time and resources on.

We also take special note of those UGZ that appear to go into outburst from standstill. When one of these stars enters a standstill we will be asking for intensive coverage until the star either goes into quiescence or outburst.

Activity is tracked in near real time as observations come in from AAVSO MyNewsFlash, BAAVSS-Alert, CVnet-Outburst, VSObs-share, and VSNET-outburst email notifications on the Activity at a Glance portion of the section home page:  https://sites.google.com/site/aavsocvsection/Home.

## 5. Early results

The Z CamPaign was launched on September 25, 2009. From increased coverage of some stars and a thorough analysis of the AAVSO light curves we have positively confirmed ten UGZ systems: RX And, TZ Per, Z Cam, AT Cnc, SY Cnc, AH Her, UZ Ser, EM Cyg, VW Vul, and HX Peg. Most of these were known or suspected UGZ.

We have also been able to identify several potential Z Cam imposters. There is no evidence of standstills in their AAVSO light curves going back decades, in most cases. Included in this group are: TW Tri, KT Per, BI Ori, CN Ori, SV CMi, and AB Dra. Some of these stars have been erroneously classified as UGZ for decades in major variable star catalogues.

V344 Ori and V391 Lyr have uncharacteristic long outburst cycles of hundreds of days, and V1363 Cyg is an unusual, unique star, but its light variations are not typical of a UGZ. None of these is a Z Cam. FY Vul has an outburst cycle between 30 and 50 days, but also shows some quasi-periodic variation on shorter time scales, perhaps 15 to 20 days. The amplitude of variation is rather small for a UGZ type dwarf nova. It has been suggested that this star and V1101 Aql may represent a previously unrecognized group of low-amplitude dwarf novae (Kato *et al.* 1999).

As a result of nearly continuous coverage by CCD observers we have uncovered a previously unknown behavior in IW And: repeated outbursts to standstills followed by another outburst and then a rapid fade to quiescence. This whole process then repeats (see Figure 1). Even more interesting is we have found the same unusual behavior in V513 Cas (see Figure 2).

## 6. Conclusion

Depending on which catalogue you reference, there are only 30 to 40 Z Cam dwarf novae. If a significant percentage of suspected Z Cams eventually proves not to be Z Cam, the remaining few represent a fairly rare and unique class of stars worthy of further investigation. Z Cam stars are rather ignored for the most part by amateur and professional alike. This leaves the door to discovery open for those patient and persistent enough to devote time and energy to long-term monitoring of this unique class of cataclysmic variable. We plan to continue this campaign through 2011, modifying the targets list as new information becomes available.

**7. Acknowledgements**

We acknowledge with thanks the variable star observations from the AAVSO International Database contributed by observers worldwide and used in this research. Several individual observers have been of particular help to this campaign: Gary Poyner, Richard Sabo, George Sjoberg, Tim Crawford, Kenneth Menzies, David Boyd, Bart Staels, Ken Mogul, Keith Graham, and Bill Goff.

| Name | R.A. (2000) h m s | Dec. (2000) ° ' " | Con- stellation | Type | Outburst Cycle (d) | Range |
|---|---|---|---|---|---|---|
| V513 Cas | 00 18 14.90 | +66 18 14.0 | Cas | UGZ: | — | 15.5–<17.2 p |
| IW And | 01 01 08.90 | +43 23 26.0 | And | UGZ | — | 14.2–17.4 p |
| RX And | 01 04 35.50 | +41 17 58.0 | And | UGZ | (14) | 10.3–14 V |
| TW Tri | 01 36 37.00 | +32 00 40.0 | Tri | UGZ: | (28) | 13.3–17.0 p |
| KT Per | 01 37 08.50 | +50 57 20.0 | Per | UGZ+ZZ | (26) | 11.5–15.39 V |
| TZ Per | 02 13 50.90 | +58 22 53.0 | Per | UGZ | (17) | 12–15.6 V |
| PY Per | 02 50 00.10 | +37 39 23.0 | Per | UGZ | — | 13.8–16.5 p |
| BI Ori | 05 23 51.80 | +01 00 30.0 | Ori | UGZ | (20.5) | 13.2–16.7 p |
| CN Ori | 05 52 07.80 | –05 25 01.0 | Ori | UGZ | (15.85) | 11–16.2 V |
| V344 Ori | 06 15 19.00 | +15 31 00.0 | Ori | UGZ: | — | 14.2–17.5: p |
| WZ CMa | 07 18 49.20 | –27 07 43.0 | CMa | UGZ: | (27.1) | 14.5–<16.0 p |
| SV CMi | 07 31 08.40 | +05 58 49.0 | CMi | UGZ | (16:) | 13.0–16.3 p |
| BX Pup | 07 54 15.60 | –24 19 36.0 | Pup | UGZ | (18) | 13.76–16 V |
| Z Cam | 08 25 13.20 | +73 06 39.0 | Cam | UGZ | (22) | 10–14.5 V |
| AT Cnc | 08 28 36.90 | +25 20 03.0 | Cnc | UGZ | (14) | 12.3–14.6 p |
| SY Cnc | 09 01 03.32 | +17 53 56.0 | Cnc | UGZ | — | 10.6–14.0 p |
| AH Her | 16 44 10.00 | +25 15 02.0 | Her | UGZ | (19.8) | 10.9–14.7 p |
| UZ Ser | 18 11 24.90 | –14 55 34.0 | Ser | UGZ | (26.4) | 12.0–16.7 p |
| V391 Lyr | 18 21 12.00 | +38 47 44.0 | Lyr | UGZ: | (100:) | 14.0–17.0 p |
| HS 1857+7127 | 18 57 20.40 | +71 31 19.2 | Dra | UGZ+E | — | 13.9–17.2 |
| V868 Cyg | 19 29 04.40 | +28 54 26.0 | Cyg | UGZ: | (20.38) | 14.3–<17.8 p |
| V1505 Cyg | 19 29 49.00 | +28 32 54.0 | Cyg | UGZ: | — | 15.2–<17.5 p |
| EM Cyg | 19 38 40.10 | +30 30 28.0 | Cyg | UGZ+E | — | 11.9–14.4 p |
| FY Vul | 19 41 40.00 | +21 45 59.0 | Vul | UGZ:/NL | — | 13.4–15.33 B |
| AB Dra | 19 49 06.40 | +77 44 23.0 | Dra | UGZ | (13.4) | 11–15.3 V |
| V1363 Cyg | 20 06 11.60 | +33 42 38.0 | Cyg | UGZ:/UGSU: | — | 13.0–<17.6 p |
| VW Vul | 20 57 45.10 | +25 30 26.0 | Vul | UGZ | (30) | 13.1–16.27 B |
| V1404 Cyg | 21 57 16.40 | +52 12 00.0 | Cyg | UGZ: | (19.15) | 15.7–<17.7 p |
| MN Lac | 22 23 04.60 | +52 40 58.0 | Lac | UGZ: | — | 15.1–<18.0 p |
| HX Peg | 23 40 23.70 | +12 37 42.0 | Peg | UGZ | — | 12.9–16.62 V |

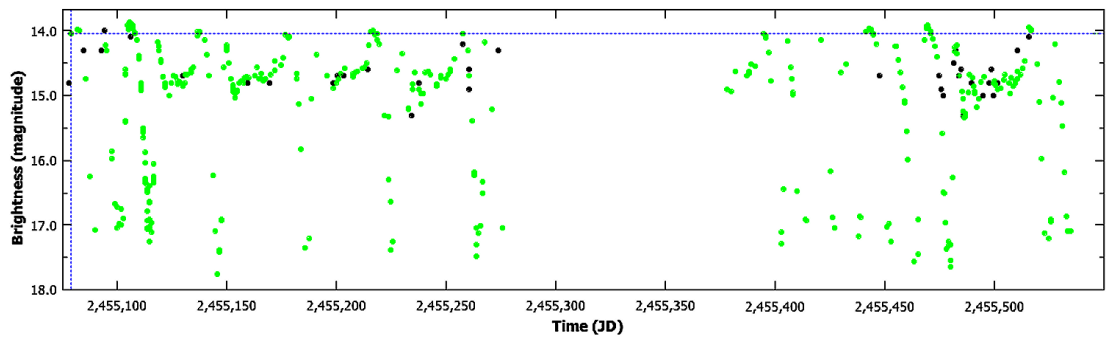

Figure 1. Light curve for IW And, September 1, 2009–March 1, 2010, showing quasi-periodic eclipse-like fadings. Visual observations (dark points), and Johnson *V* observations (lighter points) are shown.

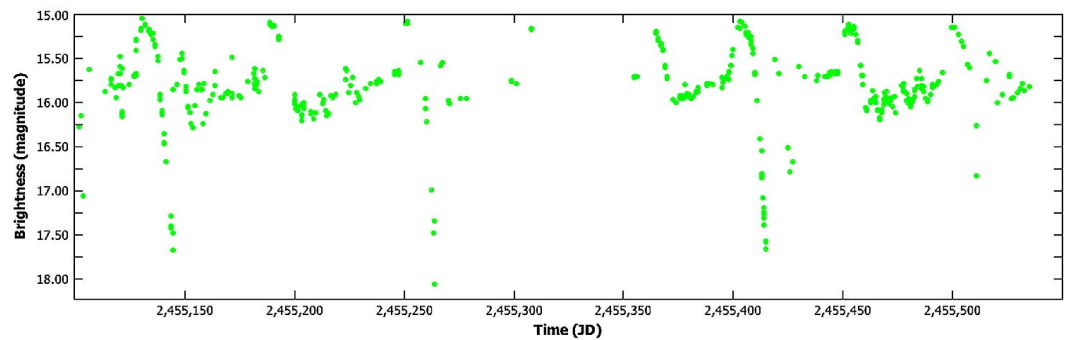

Figure 2. Light curve for V513 Cas, September 1, 2009–September 1, 2010, showing features similar to IW And. Johnson *V* observations are shown.